\documentclass [11pt]{article}
\usepackage{amsfonts}

\newtheorem{theorem}{Theorem}
\def\Z{{\mathbb Z}}
\def\R{{\mathbb R}}
\def\Ker{\mathrm{Ker}\,}
\def\Im{\mathrm{Im}\,}
\def\dim{\mathrm{dim}\,}
\def\rank{\mathrm{rank}\,}
\def\const{\mathrm{const}\,}

\begin{document}

\title{On a numerical algorithm for computing topological characteristics of three-dimensional bodies
\thanks{This work is supported by RFBR (grant 11-01-12106-ofi-m-2011) and the President of Russia grants
(MD-249.2011.1 and NSh-544.2012.1 ).}}
\author{Ya.V. Bazaikin
\thanks{Sobolev Institute of Mathematics, 630090 Novosibirsk, Russia; e-mail: bazaikin@math.nsc.ru}
\and
I.A. Taimanov
\thanks{Sobolev Institute of Mathematics, 630090 Novosibirsk, Russia; e-mail: taimanov@math.nsc.ru}}
\date{}

\maketitle

\begin{abstract}

We present an algorithm for computing the main topological charac\-te\-ris\-tics of three-dimensional bodies.
The algorithm is based on a discretization of Morse~theory and uses discrete analogs of smooth functions
with only nondegenerate (Morse) and the simplest degenerate critical points.

\textbf{Keywords:} computational topology, cubic complexes, Betti numbers, Morse theory
\end{abstract}

\section[]{Introduction}

The problem of computing topological characteristics of geometrical objects of complicated
shape appears in many applications. This led to the rise of computational topology
which is fast developing now  \cite{1,2,3}. Usually a problem of computational topology
reduces to effective computation of Betti numbers, i.e. the ranks of homology groups.

One of the most typical approaches to computing homology groups is based on Morse theory
\cite{4,5} which has to be modified for studying objects that are defined by discrete data. 
Discrete approaches
to the Morse theory were systematically studied in \cite{1} and \cite{6} and various discrete
realizations of Morse theory were considered in \cite{7,8,9,10}.

In this article we propose a method for computing the Betti numbers
of three-dimensional bodies. Therewith the bodies are defined as unions of unit cubes
with integer-valued coordinate of vertices in the three-dimensional Euclidean space.
The main feature of the approach is

\begin{itemize}
\item
{\sl the use of discrete analogs of smooth functions (of two variables)
which in addition to standard nondegenerate (Morse) critical
points have the simplest degenerate critical points of the type of the ``monkey saddle''
in the case under study.}
\end{itemize}

The scheme of this approach is as follows. Let $M$ be a three-dimensional body with a boundary
in the three-dimensional Euclidean space. We consider the critical points of a differentiable (smooth) function
$f$ on $M$, i.e. the points at which the gradient of $f$ vanishes. The values of $f$ at these
points are called {\it critical}.
When the value of $f$ passes through a critical value the topology of the excursion set
$\{f \leq \const\}$ changes. By using critical points we construct a chain complex
$C_\ast = C_0 \oplus C_1 \oplus C_2 $ which is a graded commutative group.
To the critical points there correspond the basic elements (generators) of the complex
$C_\ast $ as of a vector space over the commutative field
$\Z_2$ which consists of two elements.
To a nondegenerate critical point of index $i$ there corresponds a generator of
$C_i$ and to a critical point of the ``monkey saddle'' type there correspond two
generators of $C_1 $. The gradient flow of $f$ defines the boundary operators which are
homomorphisms of commutative groups $\partial _i :C_i \to C_{i - 1}$.
The final part of this algorithm is standard: since the kernel
$\Ker \partial$ of a homomorphism $\partial: C_\ast \to C_\ast $ lies in
its image $\Im \partial$   the homology groups
(with coefficients in $\Z_2)$ are correctly defined as the quotient spaces
$$
H_i = \frac{\Ker (\partial _i: C_i \to C_{i - 1} )}{\Im(\partial _{i + 1}:
C_{i + 1} \to C_i )}.
$$
The dimension of the $i$-th homology group as of a vector space over $\Z_2$
is called the $i$-th Betti number of $M$ and it is denoted as
$b_i $. The Betti numbers $b_0$, $b_1$, and $b_2$ which are interpreted
as the number of connected components, the number of linearly independent one-dimensional cycles
and the number of holes in $M$ and their alternated sum
$\chi = b_0 - b_1 + b_2 $ are the main topological invariants of $M$.

Clearly, we consider a discrete version of this algorithm. Note that since
a body is three-dimensional it is enough to restrict consideration to homology with $\Z_2$ coefficients
for a complete description of homological characteristics of $M$.
Also we may use the Alexander duality and compute only the homo\-mor\-phism
$\partial _1 :C_1 \to C_0 $. We will discuss that here in morein detail.

Since we consider only bounded bodies, every such a body is assumed to be
embedded into a cube $K$ in the three-dimensional space.
The algorithm of computing the matrices of boundary operators
has complexity $O(n)$ where $n$ is the number of vertices, points with
integer-valued coordinates, in $K$, and the complexity of computing  Betti numbers is equal
to $O(n_c^3)$, where $n_c$ is the number of critical points of $f$.

{\sc Remark 1.} There exist infinitely many different topological types of
critical points of analytic functions of $n$ ($n \geq 2$) variables and
their clas\-si\-fi\-cation is
very complicated. However there are finitely many Morse singularities and topological surgeries of
corresponding level surfaces are completely well described.
Therefore in computational topology it was supposed to use only the functions with Morse
singularities for computing topological characteristics (see the scheme below).
However it appears that after discretization the classification of critical
points becomes finite which is quite natural and by describing all topological surgeries
corresponding to non-Morse singularities one may use more general functions
for realization of the scheme mentioned above and we demonstrate that in
the three-dimensional case.

{\sc Remark 2.} There are algorithms for computing  Betti numbers
with complexity asymptotically linear in $n$,
\footnote{The complexity of effective algorithms
of distinguishing connected components is equal to
$O(n\alpha (n))$ where $\alpha (n)$ is a slowly growing function which is inverse to the
the Ackermann function.}
which substantially use the specifics of the three-dimensional case (see, for instance,
\cite{11}). In applications \cite{12} in which our algorithm was used the number
$n_c $ of critical points is small as compared with $n$ and that gives a practical advantage.
Moreover the proposed method may be generalized to higher dimensions.

\section[]{Critical points of the diagonal function on a Eu\-cli\-dean cube}

By an elementary interval $I \subset \R$ we mean the set $I
= [l,l + 1]$ for some $l \in \Z$. We analogously define the elementary square
$Q = I_1 \times I_2 \subset \R^2$ and the elementary cube
$C = I_1 \times I_2 \times I_3 \subset \R^3$, where $I_k$ are elementary intervals.

We consider three-dimensional bodies $M$ formed by elementary cubes and lying inside the bounded domain
$$
K = [0,N]\times [0,N]\times [0,N] \subset \R^3,
$$
for some natural number $N$.

In fact we compute homology not of $M$ but of the body
$\widetilde {M}$ which is constructed from $M$ by unstacking joint vertices and boundary edges
for such elementary cubes, in $M$, which intersect only at these joint vertices and edges.

We make to remarks:

1) we consider $\widetilde {M}$ due to applications of the algorithm;
\footnote{ In \cite{12} this algorithm is used for evaluation of topological
characteristics of oil reservoirs. Such a reservoir is modeled by a three-dimensional body that
is a union of three-dimensional cells. Therewith it is assumed that a passing of oil from one
cell to another is possible only though the joint two-dimensional face, and there is no passing
through a joint vertex or a joint edge. To this end, we have to make a preliminary processing of $M$.}

2) the body $\widetilde{M}$ itself, or more precisely a topologically equivalent body,
may be realized by unions of elementary cubes in a cube $K^\prime $ of a larger size,
as shown by the following construction.
Let us consider the cube
$$
K^\prime = [0,3N]\times [0,3N]\times [0,3N],
$$
and the linear map $P(x_1 ,x_2 ,x_3 ) = (3x_1 ,3x_2 ,3x_3 )$ from $K$ to
$K^\prime $.
We put $M^\prime = P(M)$, i.e.. $M^\prime $ is a subdivision of $M$
such that every elementary edges splits into three parts and
every elementary cube splits into $9$ parts. Then we remove
from $m^\prime$ all elementary cubes such that they have a joint face with
$\partial M^\prime$. We obtain the new cubical set $\widetilde {M}$. In the sequel we assume that
$M$ is obtained, if need be by such a surgery.

Let us take for a function $f$ on $K$ the ``diagonal''
function
$$
f(x_1 ,x_2 ,x_3 ) = x_1 + x_2 + x_3 .
$$
As usual, we put
$$
M^a = \{\bar {x} \in M\vert f(\bar {x}) \le a\}.
$$
It is clear that if
$a_1 < a_2 $ and the set $f^{ - 1}([a_1 ,a_2 ]) \cap M$
does not contain points with integer-valued coordinates, then
$M^{a_1 }$ and $M^{a_2 }$
are topologically equivalent (see, for instance, \cite{4}). Therefore, under a continuous variation
of $a$ from $0$ to $3N$ the topological type of $M^a$ changes only when $a$ passes through a value of the form
$a = f(x_1 ,x_2 ,x_3 )$ where $x_i \in \Z$.

Let $k_1 $, $k_2 $, $k_3 $ be a triple of integer numbers such that $v = (k_1
,k_2 ,k_3 ) \in K$. We put
$$
N(v) = \{\bar {x} \in M\vert \vert x_i - k_i \vert \le 1,i = 1,2,3\}.
$$
Hence $N(v)$ consists of eight elementary cubes surrounding $v$ and lying in $M$. For $a = f(v)$ and $\vert b -
a\vert < \frac{1}{6}$ we put
$$
N_v M^b = N(v) \cap M^b.
$$
We say that $v \in M$ is a critical point of a function $f$ if for sufficiently small positive
$\varepsilon < \frac{1}{6}$ the sets $N_v M^{a -
\varepsilon }$ and $N_v M^{a + \varepsilon }$ are not topologically equivalent.

For distinguishing the critical point type we code the cubes from $N(v)$ as follows. This neighborhood
consists of eight elementary cubes and with respect to the central integer-valued point every cube may be
defined by changes of the coordinates whose values are either greater or less inside the cube than at $v$.
Hence each cube is coded by a triple of signs
``$\pm \pm \pm $''. Every such a cube may lie or not lie in $M$. Hence we may correspond to
$N(v)$ the pair of matrices
$$
\left( {{\begin{array}{*{20}c}
 {t_{ - \, - \, - } } \hfill & {t_{ + \, - \, - } } \hfill \\
 {t_{ - \, + \, - } } \hfill & {t_{ + \, + \, - } } \hfill \\
\end{array} }} \right) \ \ \ \ \left( {{\begin{array}{*{20}c}
 {t_{ - \, - \, + } } \hfill & {t_{ +\,  -\, + } } \hfill \\
 {t_{ - \, + \, + } } \hfill & {t_{ + \, + \, + } } \hfill \\
\end{array} }} \right),
$$
where
$t_{\ast \ast \ast } =1$ if the cube lies in $M$ and $t_{\ast \ast \ast}=0$ otherwise.

We say that a critical point $v \in M$ is nondegenerate if $N_v M^{a +
\varepsilon }$ topologi\-cal\-ly looks like the body $N_v M^{a - \varepsilon }$ with a handle of index $i, i=0,1,2$,
glued to its boundary, i.e. one of the three possibilities holds:

1) (a handle of index $0$) $N_v M^{a - \varepsilon }$ is empty and there appears a new connected component
$N_v M^{a + \varepsilon }$;

2) (a handle of index $1$) neighborhoods of two points on the boundary of $N_v M^{a -
\varepsilon }$ are connected by a thickened interval;

3) (a handle of index $2$) to a tubular (ribbon) neighborhood of a circle
in the boundary of $N_v M^{a - \varepsilon }$ it is glued a thickened two-dimensional disc
therewith the boundary of disc is glued to the circle.

We call a critical point {\it degenerate} otherwise. The number $i$
is called the index of a critical point.

The following theorem is proved straightforwardly by the enumerating com\-bi\-natorial types of critical points.

\begin{theorem}
All possible type of critical points are given in Table~\ref{tab1}.

The types 1--14 correspond to nondegenerate critical points. The last type correspond to
the ``monkey saddle'' which in the smooth case if given by a critical point  $(x,y) = (0,0)$
of the function $f(x,y) = x^3 - xy^2$. The passing through this value results in gluing
two handles of index $1$.

\end{theorem}

\begin{table}[htbp]
\caption{Types of critical points}
\begin{tabular}
{|p{69pt}|p{120pt}|p{69pt}|}
\hline
№&
&
index \\
\hline 1.
&
$\left( {{\begin{array}{*{20}c}
 0 \hfill & 0 \hfill \\
 0 \hfill & 0 \hfill \\
\end{array} }} \right)\left( {{\begin{array}{*{20}c}
 0 \hfill & 0 \hfill \\
 0 \hfill & 1 \hfill \\
\end{array} }} \right)$&
0 \\
\hline 2.
&
$\left( {{\begin{array}{*{20}c}
 0 \hfill & 1 \hfill \\
 1 \hfill & 1 \hfill \\
\end{array} }} \right)\left( {{\begin{array}{*{20}c}
 1 \hfill & 1 \hfill \\
 1 \hfill & 1 \hfill \\
\end{array} }} \right)$&
2 \\
\hline 3.
&
$\left( {{\begin{array}{*{20}c}
 0 \hfill & 0 \hfill \\
 0 \hfill & 0 \hfill \\
\end{array} }} \right)\left( {{\begin{array}{*{20}c}
 0 \hfill & 1 \hfill \\
 1 \hfill & 1 \hfill \\
\end{array} }} \right)$&
1 \\
\hline 4.
&
$\left( {{\begin{array}{*{20}c}
 0 \hfill & 0 \hfill \\
 0 \hfill & 1 \hfill \\
\end{array} }} \right)\left( {{\begin{array}{*{20}c}
 0 \hfill & 1 \hfill \\
 0 \hfill & 1 \hfill \\
\end{array} }} \right)$&
1 \\
\hline 5.
&
$\left( {{\begin{array}{*{20}c}
 0 \hfill & 0 \hfill \\
 0 \hfill & 1 \hfill \\
\end{array} }} \right)\left( {{\begin{array}{*{20}c}
 0 \hfill & 0 \hfill \\
 1 \hfill & 1 \hfill \\
\end{array} }} \right)$&
1 \\
\hline 6.
&
$\left( {{\begin{array}{*{20}c}
 0 \hfill & 0 \hfill \\
 0 \hfill & 1 \hfill \\
\end{array} }} \right)\left( {{\begin{array}{*{20}c}
 1 \hfill & 1 \hfill \\
 1 \hfill & 1 \hfill \\
\end{array} }} \right)$&
1 \\
\hline 7.
&
$\left( {{\begin{array}{*{20}c}
 0 \hfill & 1 \hfill \\
 0 \hfill & 1 \hfill \\
\end{array} }} \right)\left( {{\begin{array}{*{20}c}
 0 \hfill & 1 \hfill \\
 1 \hfill & 1 \hfill \\
\end{array} }} \right)$&
1 \\
\hline 8.
&
$\left( {{\begin{array}{*{20}c}
 0 \hfill & 0 \hfill \\
 1 \hfill & 1 \hfill \\
\end{array} }} \right)\left( {{\begin{array}{*{20}c}
 0 \hfill & 1 \hfill \\
 1 \hfill & 1 \hfill \\
\end{array} }} \right)$&
1 \\
\hline 9.
&
$\left( {{\begin{array}{*{20}c}
 0 \hfill & 1 \hfill \\
 0 \hfill & 0 \hfill \\
\end{array} }} \right)\left( {{\begin{array}{*{20}c}
 0 \hfill & 1 \hfill \\
 1 \hfill & 1 \hfill \\
\end{array} }} \right)$&
1 \\
\hline 10.
&
$\left( {{\begin{array}{*{20}c}
 0 \hfill & 0 \hfill \\
 1 \hfill & 1 \hfill \\
\end{array} }} \right)\left( {{\begin{array}{*{20}c}
 0 \hfill & 1 \hfill \\
 0 \hfill & 1 \hfill \\
\end{array} }} \right)$&
1 \\
\hline 11.
&
$\left( {{\begin{array}{*{20}c}
 0 \hfill & 0 \hfill \\
 0 \hfill & 1 \hfill \\
\end{array} }} \right)\left( {{\begin{array}{*{20}c}
 1 \hfill & 0 \hfill \\
 1 \hfill & 1 \hfill \\
\end{array} }} \right)$&
1 \\
\hline 12.
&
$\left( {{\begin{array}{*{20}c}
 0 \hfill & 0 \hfill \\
 1 \hfill & 0 \hfill \\
\end{array} }} \right)\left( {{\begin{array}{*{20}c}
 0 \hfill & 1 \hfill \\
 1 \hfill & 1 \hfill \\
\end{array} }} \right)$&
1 \\
\hline 13.
&
$\left( {{\begin{array}{*{20}c}
 0 \hfill & 0 \hfill \\
 0 \hfill & 1 \hfill \\
\end{array} }} \right)\left( {{\begin{array}{*{20}c}
 1 \hfill & 1 \hfill \\
 0 \hfill & 1 \hfill \\
\end{array} }} \right)$&
1 \\
\hline 14.
&
$\left( {{\begin{array}{*{20}c}
 0 \hfill & 1 \hfill \\
 0 \hfill & 1 \hfill \\
\end{array} }} \right)\left( {{\begin{array}{*{20}c}
 0 \hfill & 0 \hfill \\
 1 \hfill & 1 \hfill \\
\end{array} }} \right)$&
1 \\
\hline 15.
&
$\left( {{\begin{array}{*{20}c}
 0 \hfill & 0 \hfill \\
 0 \hfill & 1 \hfill \\
\end{array} }} \right)\left( {{\begin{array}{*{20}c}
 0 \hfill & 1 \hfill \\
 1 \hfill & 1 \hfill \\
\end{array} }} \right)$&
the monkey saddle \\
\hline
\end{tabular}
\label{tab1}
\end{table}

\vskip0.2cm

Let $M_i $, $i = 0,1,2$, be the sets of critical points of index $i$. Every monkey saddle $v$ 
contributes to $M_1$ the two points:
the point $v$ itself and its formal ``double'' $v^\prime$. 
For every $i = 0,1,2$ let us consider the vector space $C_i $ over $\Z_2$ with generators from $M_i $.
To construct a chain complex we have to define the linear boundary operators
$\partial _1 :C_1 \to C_0 $ and $\partial _2:C_2 \to C_1$.

\section[]{The gradient flow and the boundary operators}

The gradient flow enables us to describe a gluing of handles and
that allows us to compute the boundary operators
in the chain complex $C_*$. By a trajectory of
the gradient flow we mean a sequence of vertices of $M$ which we may pass going down along edges.
The difference from the gradient flow of a smooth function
consists in an indeterminacy, i.e., from every vertex we may descend into not a unique
vertex that lies below. Since we use the Alexander duality, it is enough
to define the gradient flow from critical points of index $1$ to critical points of index $0$.

Let us recall that we compute homology with coefficients in the filed
$\Z_2 $ which consist of the two elements $0$ and $1$ meeting the summation rules:
$0+0=1+1=0, 0+1=1+0=1$.

By a tangent vector $\xi = (v,e) \in TM$ we mean a pair consisting of
a vertex $v$ and an edge $e \ni v$ which lie in the boundary of
$M$. We say that $\xi $ is increasing if another end of $e$
lies at a higher level of $f$; otherwise, we say that $\xi$ is a decreasing vector.
Moreover on the set of increasing (or decreasing) tangent vectors at a given point we fix
a natural order that is induced by ordering of the coordinate directions.

Let $v$ be a critical point of $f$. If $v$ has index $0$, then exactly three tangent vectors
are drawn from $v$ and all of them are increasing. If $v$ is a nondegenerate point of index $1$, then
to every edge of a decreasing tangent vector we assign some label $S$ that is equal to 
$\pm 1$ and is as follows:
The list of all types of critical points shows that all such edges lie in the
boundary of $M$ and there are at least two of them.
If there are exactly two such edges then we mark one by $+1$ and another by $-1$.
If there are three such edges then two of them belong to an elementary face
from the boundary of $M$ and we mark these two edges by the same label, say by $+1$,
and then mark the  left edge by $-1$.

{\sc Monkey saddle.} If $v$ is a monkey saddle, then we add a fictive
critical point $v^\prime $, which lies above $v$ with respect to the level surface of $f$, and also
add a fictive edge $vv^\prime $, connecting $v$ and $v^\prime $. We put that
$(v^\prime ,(1,0,0))$, $(v^\prime ,(0,1,0))$,
$(v,(0,0,1))$ and $(v,vv^\prime )$ are increasing vectors and
$(v^\prime ,(0,0,-1))$, $(v^\prime ,v^\prime v)$, $(v^\prime,(-1,0,0))$, and $(v^\prime,(0,
- 1,0))$ are decreasing vectors. We put that the end of every non-zero vector
$(v^\prime,(\alpha,\beta,\gamma))$ coincides with the end of
$(v, (\alpha, \beta, \gamma))$, $\alpha,\beta, \gamma=-1,0,1$.

Let $v$ be a critical point of index $0$, and $v^\prime$ be a critical point of index~$1$.
A pair of tangent vectors $\xi = (v,e)$ and $\xi ^\prime =
(v^\prime ,e^\prime )$ is called a \textit{gradient} pair, if $\xi$ is an increasing vector,
$\xi^\prime $ is a decreasing vector, and the following conditions hold:
there exists a sequence of increasing tangent vectors
$(e_i ,v_i )$, $i = 0,\ldots ,k$ such that $e_0 = e$, $v_0 = v$, the edge $e_i $
connects $v_i $ and $v_{i + 1} $, $e_k = e^\prime $, and for every vertex $v_i $
the vector $(e_i, v_i)$ is greater (with respect to the order) than all other
tangent vectors which start at this vertex and lie in $M$.

Exactly in this case we will write that $\xi < \xi^\prime $.
We mean by the gradient flow of $f$ on $M$ the family of gradient pairs of tangent vectors:
$$
GF(M) = \{(\xi _1 ,\xi _2 )\vert \xi _i \in TM,i = 1,2,\xi _1 < \xi _2 \}.
$$

By definition, a path in $M$ is a family of successively traversed edges from $M$.
The path formed by edges $e_0,\dots,e_k$ from the definition of a gradient pair is called gradient.

We define the boundary linear operator by the formula:
$$
\partial_1 v_1 = v_0^+ + v_0^-,\ \ \ v_1 \in M_1,
$$
where
$$
v_0^ + \in \{v \in M_0 \vert \exists \xi = (v,e), \exists \xi _1 = (v_1 ,e_1
),S(e_1 ) = + 1,\xi < \xi _1 \},
$$
$$
v_0 \in \{v \in M_0 \vert \exists \xi = (v,e), \exists \xi _1 = (v_1 ,e_1
),S(e_1 ) = - 1,\xi < \xi _1 \}.
$$
We recall that $M_0$ and $M_1$ are the sets of critical points of indices $0$ and $1$.

Let us remark that this definition is correct because
from every critical point of index $1$ there are drawn two decreasing tangent vectors
labeled by different signs. Therewith we may go down until we achieve (in finitely many steps)
a critical point of index $0$. Moreover
$v_0^+$ and $v_0^-$ are uniquely defined because we always choose the highest tangent vector.
If we consider two gradient paths $\gamma _1 $ and $\gamma _1^\prime $
going down to $v_0^+ $ and $v_0^-$, then together these paths form a path whose neighborhood
forms a handle of index $1$ attached when the value of $f$ passes through $f(v_1)$ (see \S 2).

Let us consider the three-dimensional body $M^\prime = K \setminus M$, the complement to $M$, and
a function $h = - f$ on $M^\prime$. Clearly, $M_i^\prime = M_{2 - i}$, $i =
1,2$; $M_0 ^\prime = M_2 \cup \{p_0 \}$, where $p_0 $ is the corner point of $K$,
where $h$ achieves its minimal value.
Let $C_i ^\prime $ be the vector space over $\Z_2$ generated by the elements of $M_i^\prime$.
We have the isomorphisms $C_i^\prime = C_{2 - i}$, $i=1,2$, and $C_0 ^\prime =
C_2 \oplus \Z_2$, where the $Z_2$ component is generated by $p_0$.
By applying the construction of $\partial_1$ to $h$ we obtain the boundary operator
$\partial_{1\,h}: C_1^\prime \to C_0^\prime$.
Let us consider the dual spaces
$C_i^{\prime \ast}$, $i=0,1,2$, consisting of $\Z_2$-valued linear functions on $C_i^\prime$.
There are isomorphisms $C_i^{\prime \ast}=C_i^\prime$, $i=0,1,2$, which assign to every
$Z_2$-function on $C_i^\prime$ the formal sum of points from $M_i^\prime$, at which the
function does not vanish.
The operator $\partial_{1\,h}$ defines the adjoint operator
$$
\delta_0=\partial_{1 h}^*: C_0^{\prime \ast} = C_2 \oplus \Z_2 \rightarrow C_1^{\prime \ast} = C_1.
$$
We put
$$
\partial_2=\delta_0\big|_{C_2}: C_2 \rightarrow C_1.
$$

\begin{theorem}
The homology groups of the chain complex $C_\ast = (C_i ,\partial _i )$ are isomorphic to
the homology groups of $M$ with coefficients in $\Z_2: H_i(C_*)=H_i(M;\Z_2)$, $i=0,1,2$.
\end{theorem}

{\it Proof.}
For every critical point of index $1$ there exist two paths such that they go down to critical points of index $0$
in the directions labeled by different marks. This pair of gradient paths is combined into the path that
connects two critical points of index $0$ and hence the formal sum of these points belongs to
the image of $\partial_1$. By using the standard reasonings of Morse theory, it is easily proved that if two
critical points of index $0$ lie in the same connected component of $M$ then they are
connected by the chain of such pairs of gradient paths.
Since every connected component of $M$ contains at least one critical point of index $0$, we have
$H_0 (M) = C_0 / \Im \partial_1$. Analogously it is shown that $H^0(M^\prime ) = \Ker \delta_0$.
Then we have $\widetilde{H}^0(M^\prime) = \Ker \partial_2$. Here we recall that
the reduced homology groups $\widetilde{H}_i$ meet the equalities
$\widetilde{H}_i(M)=H_i(M)$ for $i>0$ and  $H_0(M)=\widetilde{H}_0(M) \oplus \Z_2$;
the analogous equalities are valid for the reduced cohomology groups.

Let us recall the Alexander duality theorem
(see, for instance, \cite{13}):

{\sl Let $A \subset \R^n$ be a compact polyhedron. Then we have the isomorphism:}
$$
\widetilde{H}_q (\R^n \setminus A;\Z_2) \approx \widetilde{H}^{n - q - 1}(A;\Z_2).
$$

By definitions and duality, we have
$H_2 (M) = \widetilde{H}_2 (M) = \widetilde{H}^0 (M^\prime ) =
\Ker(\partial_2 )$. Hence the $0$-th and the $2$-nd homology groups of
$C_\ast $ are isomorphic to the corresponding homology groups of $M$.

The Euler characteristic $\chi(M)$ of $M$ is equal to the alternated sum of the numbers of cells of
any cellular decomposition of a deformation retract of $M$ and, in particular, to
$\chi(C_\ast)$, and it is also equal to the alternated sum of the Betti numbers, i.e. to
$b_0 - b_1 + b_2$. Since $\chi(C_\ast) = b_0(C_\ast) - b_1(C_\ast)+b_2(C_\ast)$, we have
$$
\chi(M) = \chi(C_\ast).
$$
It suffices to notice that $b_1 (M) = b_0 (M) + b_2 (M) - \chi (M) =
b_0 (C_\ast ) + b_2 (C_\ast ) - \chi (C_\ast ) = b_1 (C_\ast )$, and, since we consider homology
with coefficients in a field, $H_1 (M) = H_1
(C_\ast )$. The theorem is proved.

\section{An algorithm for computing the boundary ope\-ra\-tors}

The approach described above may be used for constructing an algorithm as follows:
Initially we have some array $M = M[i,j,k]$, where $i,j,k = 1,\ldots ,N$, whose entries are equal to
$0$ or $1$. Therewith the three-dimensional body $M$ is formed by all elementary cubes
$[i,i + 1]\times [j,j + 1]\times [k,k + 1]$ with $M[i,j,k] = 1$.

The preprocessing consists in the following. We stretch the array $M$ in every direction, i.e. replace it
by the array $M^\prime$ such that
$M[i,j,k]=M^\prime[3i+i_1, 3j+j_1, 3k+k_1]$, $i_1,j_1,k_1=0,1,2$, $i,j,k = 1,\ldots N$.
Then we assign zero values to such elements $M^\prime [i,j,k]$ that correspond to elementary cubes
$[i,i+1]\times [j,j+1]\times [k,k+1]$ intersecting the boundary of $M^\prime$ over a two-dimensional face.

At the input of the algorithm we have a preprocessed array
$M$. At the output we obtain the lists $C0$ and $C1$ of critical points of indices $0$ and $1$
(as it is described above every monkey saddle contributes into $C1$ two elements; these lists give
bases for the groups $C_0$ and $C_1 )$ and the matrix $D = D[i,j]$ that defines the boundary operator
$\partial _1 $.

To every vertex $v$ we correspond a formally adjoint vertex $v^\prime $. Let us consider the
function $Ind$ which assigns to every vertex $v \in M$ either its index, if
the vertex is a critical point of index $0$ or $1$; either $-1$, if it is a monkey saddle; either
$-2$ otherwise.
The list  $C0$ contains the critical points of index $0$, and the list $C1$ consists of the critical points of
index $1$ and the points $v$ and $v'$ which correspond to every monkey saddle $v$.
The function $Num$ returns the number of a critical point of index $i$ in $Ci$. To every vertex $v$
there corresponds $GF(v)$, the number of a certain critical point of index $0$ to which one may go down from
$v$. Finally, to every vertex $v$ the functions $Down\_1$ and $Down\_2$ return the ends of
decreasing tangent vectors starting at $v$, where $Down\_1$ corresponds to the highest vector, or the
empty values if there are no such vectors.
Therewith to the critical points of index $1$ the functions
$Down\_1$ and $Down\_2$ return vertices corresponding to decreasing vectors with different signs.
If $v$ is a monkey saddle, the we assume that $Down\_1(v)$ and
$Down\_2(v)$ are the ends of vectors $(-1,0,0)$ and $(0,-1,0)$ attached to $v$.
For the adjoint vertex we put $Down\_1(v^\prime ) = v$ and $Down\_2(v^\prime )$ is the end of vector
$(0,0, - 1)$ attached to $v$. The description of the algorithm is given in Table \ref{tab2}.

\begin{table}{}
\caption{Algorithm}

\textbf{Input:} the preprocessed array $M$

\textbf{Output:} the lists of critical points $C0$ and $C1$, the boundary operator matrix $D$

\begin{tabular}{ll}

1:  & $D: = 0$, $C0=\emptyset$, $C1=\emptyset$ \\

2: & \textbf{for} for all integer-valued vertices $v \in M$ \textbf{do} \\

3: & \hskip10mm \textbf{if }$Ind(v) = 0$ \textbf{then }  \\

4: & \hskip20mm $Add(C0,v)$ \\

5: & \hskip20mm $GF(v): = \{Num(v)\}$ \\

6: & \hskip10mm \textbf{if} $Ind(v) = 1$ \textbf{then } \\

7: & \hskip20mm $Add(C1,v)$ \\

8: & \hskip20mm $GF(v): = GF(Down\_1(v))$ \\

9: & \hskip20mm \textbf{if }$GF(Down\_1(v)) \ne GF(Down\_2(v))$ \textbf{then } \\

10: & \hskip30mm $D(Num(v),Num(GF(Down\_1(v)))): = 1$ \\

11: & \hskip30mm $D(Num(v),Num(GF(Down\_2(v)))): = 1$ \\

12: & \hskip10mm \textbf{if} $Ind(v) = - 1$ \textbf{then} \\

13: & \hskip20mm $Add(C1,v)$ \\

14: & \hskip20mm $Add(C1,v^\prime )$ \\

15: & \hskip20mm $GF(v): = GF(Down\_1(v))$ \\

16: & \hskip20mm $GF(v^\prime ): = GF(Down\_1(v))$ \\

17: & \hskip20mm \textbf{if }$GF(Down\_1(v)) \ne GF(Down\_2(v))$ \textbf{then } \\

18: & \hskip30mm $D(Num(v),Num(GF(Down\_1(v)))): = 1$ \\

19: & \hskip30mm $D(Num(v),Num(GF(Down\_2(v)))): = 1$ \\

20: & \hskip20mm \textbf{if }$GF(Down\_1(v^\prime )) \ne GF(Down\_2(v^\prime ))$ \textbf{then }  \\

21: & \hskip30mm $D(Num(v^\prime ),Num(GF(Down\_1(v^\prime )))): = 1$ \\

22: & \hskip30mm $D(Num(v^\prime ),Num(GF(Down\_2(v^\prime )))): = 1$ \\

23: & \hskip10mm \textbf{if }$Ind(v) = - 2$ \textbf{then } \\

24: & \hskip20mm $GF(v): = GF(Down\_1(v))$ \\

25: & \textbf{end}

\end{tabular}
\label{tab2}
\end{table}

Applying this algorithm twice to $M$ and its complement $M^\prime $, obtain the matrices $D1$ and $D2$
describing the operators $\partial _1$ and $\partial _2$. Afterwards the Betti numbers are found from the formulas:
$$
b_0 = \dim(C0) - \rank(D1),
$$
$$
b_1 = \dim(C1) - \rank(D1) - \rank(D2),
$$
$$
b_2 = \dim(C2) - \rank(D2).
$$
The complexity of the algorithm for constructing the matrices of the boundary operators, including the preprocessing of data, is equal to $O(n)$ where $n = N^3$ is the number of vertices in $K$. For a subsequent computation of Betti numbers we have to compute the ranks of $Di$, for instance, by the Gauss method, and that needs
$O(n_c^3 )$ operations where $n_c $ is the number of critical points. Generically the relation between
$n$ and $n_c $ depends on the initial data.

\end{document}